\documentclass[aps, prd, amsmath, floats, floatfix, onecolumn, superscriptaddress, nofootinbib, 10pt]{revtex4-2}

\usepackage{amsfonts}
\usepackage{slashed}
\usepackage{physics}
\usepackage[makeroom]{cancel}

\usepackage{graphicx}
\usepackage{color}
\usepackage{latexsym}
\usepackage{slashed}
\usepackage{amsthm}
\usepackage{amssymb}
\usepackage{indentfirst} 
\definecolor{ao(english)}{rgb}{0.0, 0.5, 0.0}
\usepackage[breaklinks,pdfpagelabels]{hyperref} 
\hypersetup{colorlinks=true,linkcolor=blue,urlcolor=blue,citecolor=ao(english)}
\usepackage{mathrsfs}
\usepackage[latin1,utf8]{inputenc}

\usepackage[]{mdframed}

\newcommand{\be}{\begin{eqnarray}}
\newcommand{\ee}{\end{eqnarray}}

\begin{document}

\title{Quantum avoidance of G\"{o}del's closed timelike curves}

\author{ZHAO Zhe}
\email[]{12032049@mail.sustech.edu.cn}

\author{Leonardo Modesto}
\email[]{lmodesto@sustech.edu.cn}

\affiliation{Department of Physics, Southern University of Science and Technology, Shenzhen 518055, China}
\date{\today}

\begin{abstract}
In a large class of nonlocal as well as local higher derivative theories minimally  
coupled to the matter sector,
we investigate the exactness of two different classes of homogeneous G\"{o}del-type solutions, which may or may not allow closed time-like curves (CTC). Our analysis is limited to spacetimes solving the Einstein's EoM, thus we can not exclude the presence of other G\"{o}del-type solutions solving the EoM of local and nonlocal higher derivative theories but not the Einstein's EoM. 
%
%
%
It turns out that the homogeneous G\"{o}del spacetimes without CTC are basically exact solutions for all theories, while the metrics with CTC are not exact solutions of (super-)renormalizable local or nonlocal gravitational theories. 
Hence, the quantum renormalizability property excludes theories suffering of the G\"{o}del's causality violation. 
We also comment about nonlocal gravity non-minimally coupled to matter. In this class of theories, all the G\"{o}del's spacetimes, with or without CTC, are exact solutions at classical level. However, the quantum corrections, although perturbative, very likely spoil the exactness of such solutions.
Therefore, we can state that the G\"{o}del's Universes with CTC and the super-renormalizability are mutually exclusive.
%
%
%


%

\end{abstract}


\maketitle
\tableofcontents

\section*{Introduction}
\label{intro}
In a very inspiring and relevant paper \cite{Nascimento:2021bzb}, the authors investigated the presence of homogeneous cosmological G\"{o}del-type solutions in a quite general class of nonlocal gravitational theories. 
We here expand on the analysis in \cite{Nascimento:2021bzb} looking for a principle of mutual exclusion between unitarity (as already suggested in \cite{Nascimento:2021bzb}) and/or renormalizabilty and the presence of homogeneous G\"{o}del-type exact solutions with CTC in nonlocal gravity.
We will mainly study a classe of nonlocal theories characterized by the minimal 
coupling to gravity \cite{Koshelev:2017ebj, Koshelev:2016xqb}. However, in the end we will also comment on the same issue in a class of nonlocal gravitational theories with 
non-minimal coupling to matter \cite{nl-nm-matter}. 
%
Let us now introduce the theory and its main properties. 
Nonlocal quantum gravity has been extensively studied since $2011$ as a consistent proposal for quantum gravity in the quantum field theory framework. The minimal action for pure gravity was studied in \cite{modesto}, but subsequent further searches revealed the existence of two cornerstone papers on nonlocal quantum gravity by Krasnikov \cite{krasnikov} and Kuzmin \cite{kuzmin}. In the former paper it was proposed a tree-level unitary action with exponential non locality, while in the latter the power counting super-renormalizability \cite{Asorey:1996hz} was rigorously proved for a class of asymptotically polynomial nonlocal theories. 
In  \cite{modestoDdim} the theory was extended to any dimension and proved to be finite at quantum level in odd dimension \cite{review} (see also \cite{GianlucaLeslawReview} for a more recent review). 
In \cite{modestoLeslaw}, the theories proposed in \cite{modesto} and \cite{krasnikov, kuzmin} were extended by adding few other local operators in order to achieve the quantum finiteness in any dimension. 
The other crucial property satisfied by these theories is the perturbative unitarity which has been rigorously proved in \cite{Briscese:2018oyx, Briscese:2021mob, Liu:2022gun}. In short, the main idea is to evaluate all the loop-integral along the imaginary axis for the energy, assuming also the external energy to be purely imaginary, and, afterwards make the analytic continuation to the real physical energy. It turns out that the Cutkosky rules are the same of the local theory with which the nonlocal theory shares the perturbative spectrum, namely the asymptotic degrees of freedom.

We come now to the main topic of the paper, namely the G\"{o}del's spacetimes. 
In 1949, G\"{o}del \cite{Godel:1949ga} discovered an exact solution of the Einstein's field equations sourced by a negative cosmological constant and a pressure-free perfect fluid. 
The line element of the G\"{o}del Universe reads: 
\be
\dd{s}^2=-\qty[\dd{t}+H(x)\dd{y}]^2+D(x)\dd{y}^2+\dd{x}^2+\dd{z}^2 \, , \quad 
H(x)=e^{mx},\quad  D(x)=\frac{e^{mx}}{\sqrt2} \, , 
\label{Goedel}
\ee
and it is sourced by the following energy-momentum tensor, 
\be
T_{ab}=\rho V_aV_b, \quad V^a= (\partial_t)^a,\quad m^2=-2\Lambda_{\rm cc}=\kappa^2\rho=2\omega^2 \, , 
\ee
where $\rho$ is the constant density of matter, $V^a$ is the four-velocity of fluid, $\Lambda_{\rm cc}$ is the negative cosmological constant, $\kappa$ is related to the Newton's gravitational constant, $\omega$ is the rotation velocity of matter. 

The G\"{o}del spacetime is geodetically complete, but it presents CTC that could lead to a violation of macro-causality. The latter statement is actually debated because 
the CTC are not geodesics and can not be traveled in a finite amount of time. 
However, in this paper we will only focus on the presence of G\"{o}del's solutions in a large class of gravitational theories without investigating the geodesic motion of probe particles. 

More recently, 
in 1982, a class of homogeneous spacetimes of G\"{o}del-type \cite{Hawking:GR,ST} was proposed by M.J. Rebou\c{c}as and J. Tiomno \cite{ST-G}. 
The latter metric is homogeneous 
\cite{ISO}
and causal (no CTC) whether the parameters in the metric take specific values. This was the first proposal for a completely causal homogeneous and rotating Universe. 
In $2021$ the homogeneous G\"{o}del-type without CTC has been shown to be an exact causal solution of a special class of classical nonlocal gravitational theories \cite{Nascimento:2021bzb}. Moreover, 
in \cite{Nascimento:2021bzb} the authors did not find G\"{o}del-type solutions with CTC. 
In this paper, we generalize the result in \cite{Nascimento:2021bzb} deriving the general constrain that a local or nonlocal higher derivative theory has to satisfy whether we want the G\"{o}del-type metrics to be exact solutions of such theory. 
In this project , we will assume the G\"{o}del-type metrics to be exact solutions of the Einstein's two-derivative theory. Therefore, we can not exclude other G\"{o}del-type metrics that do not solve the Einstein's equations, but solve the full theory. 

Let us here list the issues addressed in each section of the paper.
In Sec.\ref{sec-metric}, we review the homogeneous G\"{o}del-type metrics: the classification and the conditions for the existence of CTC. 
In Sec.\ref{notation}, we introduce the notations and the implicit form of a general action for local or nonlocal gravitational theory minimally coupled to matter. 
%
%
%
In Sec.\ref{NLG}, we explicitly introduce the action, while in Sec.\ref{solution1}, we derive the equations to be solved in order to prove or disprove the presence of exact homogeneous G\"{o}del-type solutions in local or nonlocal higher derivative theories minimally coupled to matter. 
%
%
%
In Sec.\ref{solution2}, we explicitly state in compact form the equations of motion
and 
study under which restrictions the homogeneous G\"{o}del-type spacetimes (including the metrics with CTC)
are exact solutions in higher derivative or nonlocal gravitational theories.
In Sec.\ref{solutionsNoCTC} and Sec.\ref{GNoCTC}, we show that the homogeneous G\"{o}del-type spacetimes without CTC
are exact solutions of basically all local or nonlocal theories with higher derivative operators. 
%
 %
 In Sec.\ref{quantum}, we show that the conditions for unitarity and renormalizability exclude theories having  solutions with CTC. 
Finally, in section \ref{nm-gravity}, we shortly review nonlocal gravity non-minimally coupled to matter and 
show that all G\"{o}del-type spacetimes are exact solutions by construction. However, the quantum corrections very likely will exclude such solutions.

\section{G\"{o}del-type Metrics and CTC}
\label{sec-metric}
In this section, we briefly review the homogeneous G\"{o}del-type metrics and provide the condition for having CTC (for more details we invite the reader to consult \cite{ST-G}.) 


The homogeneous G\"{o}del-type metrics are defined by the following line element in cylindrical coordinates, 
\begin{equation}\label{type_godel}
    \dd[2]{s}=-[\dd{t}+H(r)\dd\theta]^2+D^{2}(r)\dd\theta^{2}+\dd{r}^2+\dd{z}^2, 
\end{equation}
where $H(r)$ and $D(r)$ are functions of the radial coordinate $r$ and satisfy the following conditions, 
\begin{eqnarray}
\label{m-omega}
\frac{H'(r)}{D(r)}=2\omega\in\mathbb{R}\backslash\{0\}\qand\frac{D''(r)}{D(r)}=m^2\in\mathbb{R},
\label{ST}
\end{eqnarray}
where the prime stays for the derivative with respect to the radial coordinate $r$. 

As discussed in \cite{ST-G}, the homogeneous G\"{o}del-type spaces can be organized in four classes
depending on the values of the parameters $\omega$ and $m^2$:
\begin{enumerate}
\item the \textit{hyperbolic class}: $m^2>0$, $\omega\neq 0$:
\begin{eqnarray}
H(r)=\frac{2\omega}{m^2}[\cosh(m \, r)-1]\qand D(r)=\frac{1}{m}\sinh(m \, r),
\end{eqnarray}
\item the \textit{trigonometric class}: $-\mu^2=m^2<0$, $\omega\neq 0$:
\begin{eqnarray}
H(r)=\frac{2\omega}{\mu^2}[1-\cos(\mu \, r)]\qand D(r)=\frac{1}{\mu}\sin(\mu \, r),
\end{eqnarray}
\item the \textit{linear class}: $m^2=0$, $\omega\neq 0$:
\begin{eqnarray}
H(r)=\omega \, r^2\qand D(r)=r \, , 
\label{linear}
\end{eqnarray}
\item the \textit{degenerate class}: $m^2\neq 0$, $\omega=0$:
\begin{eqnarray}
H(r)=0 \, .
\label{degenerate}
\end{eqnarray}
\end{enumerate}
The reader can verify that the above four classes of metrics are exact solutions of the Einstein's EoM.
As a particular example, the G\"{o}del original metric (\ref{Goedel}) corresponds to $m^2=2\omega^2>0$.

\subsection*{Closed time-like curves}
The CTC in the spacetime (\ref{type_godel}) are characterized by the following curve, 
\be
C= \Big\{ (t,r,\theta,z); \, t, r, z= \mbox{const}, \,\, \theta \in [0, 2\pi] \Big\} \, . 
\ee
Notice that for $ t, r, z = \mbox{const}$, the $\theta$ coordinate is time-like. Therefore, the curve $C$ is a CTC because $\theta$ is a periodic angular coordinate. 

It turns out that the condition for having a CTC in an homogeneous G\"{o}del-type metric is: 
%
%
\be
\boxed{
\exists \,  r_0 \,\,\, \mathrm{s.t.} \,\,\,
G(r_0)\equiv D^2 (r_0) - H^2 (r_0) < 0 \,\,\, \Longleftrightarrow  \,\, \, \boxed{4\omega^2 >m^2 > - \infty } 
} \, . 
\label{CTCcond}
\ee

It is worth noting that there are no CTC only for the degenerate class of solutions. 


\section{G\"{o}del-type solutions in local and nonlocal gravity}
\label{notation}
In order to fix the notation we here define the action and the equations of motion (EoM). The action for a general gravitational theory reads:
\be
S= \int \dd[4]{x}\sqrt{-g}\mathcal{L}_g+S_{m}[g_{\mu\nu},\Psi]
\qand 
\mathcal{L}_g = \frac{1}{2\kappa^2} \left( R - 2 \Lambda_{\rm cc} \right) +\mathcal{L} \, , 
\label{genA}
\ee
where 
$2 \kappa^2 = 16 \pi G$, $G$ is Newton's constant, 
$\Lambda_{\rm cc}$ is the cosmological constant. $\mathcal{L}$ is the Lagrangian beyond the Einstein-Hilbert one, and $S_m$ is the action for matter. Taking the variation respect to the metric $g_{\mu\nu}$, the EoM can be written as:
\begin{equation}
	\label{dQ}
	-\frac{2}{\sqrt{-g}}\fdv{S}{g_{ab}}=0
	\quad 
	\Longrightarrow
	 \quad 
	 \frac{1}{\kappa^2}\qty(G_{ab}+\Lambda_{cc}g_{ab}+Q_{ab})=
	T_{ab}, 
\end{equation}
where $Q_{ab}$ is the extra term coming from the variation of $\mathcal{L}$.
 
 We now look for G\"{o}del-type solutions in several nonlocal and local theories. 
 
\subsection{Nonlocal gravity} \label{NLG}
In order to extend the result in \cite{Nascimento:2021bzb}, in this section 
we consider the nonlocal gravitational action 
in the Ricci-Weyl basis,
\begin{eqnarray}
\label{actmod}
S = \int \dd[4]x\sqrt{-g}\left[ \frac{1}{2\kappa^2}\left( 
R-2\Lambda_{\rm{cc}} \right) +R\gamma_{0}(\square)R+R_{ab}\gamma_{2}(\square)R^{ab}+C_{abcd}\gamma_{4}(\square)C^{abcd} \right] 
%
+S_{m}[g_{ab},\Psi] \, ,
\label{ac}
\end{eqnarray}   
where 
 the analytic form factors $\gamma_i$ are infinite power series of the dimensionless d'Alembert operator $\square_{\Lambda}\equiv \square/\Lambda^{2}$, i.e.,
\begin{equation}
\gamma_{i}(\square)=\sum_{n=0}^{\infty}\gamma_{i,n}\square^{n}_{\Lambda}\qand i=\{0,2,4\},
\label{fr}
\end{equation}
$\gamma_{i,n}$ are the coefficients (dimensionless) of the power series in $\square_{\Lambda}$, $\Lambda$ is an invariant fundamental mass scale of the theory, namely in this case the non locality scale. 

Notice that the unitarity of the theory has been investigate in the Riemann-Ricci basis \cite{Koshelev:2016xqb}, hence, we will later in section (\ref{quantum}) change basis in order to make a connection between renormalizable and unitary theories with the presence of G\"{o}del solutions.

\subsection{Equations of motion with the G\"{o}del's ansatz}
\label{solution1}
In this project we do not look for new solutions, but we investigate under which conditions the G\"{o}del-type metrics of the section (\ref{sec-metric}), which solve the Einstein's EoM, 
are solutions of the general nonlocal or local higher derivative theory (\ref{genA}). 

Therefore, in order to solve the modified EoM (\ref{dQ}) the G\"{o}del's metric $\bar{g}_{ab}$ should satisfy:
\be
Q^{\alpha \beta}(\bar{g}) \equiv 0 \, \quad \forall \,\, m  , \, \omega. 
\label{Q}
\ee
The latter equation turns out to be an algebraic constraint on the space of theories, namely on the constant coefficients 
$\gamma_{i, n}$.
Indeed,
\begin{equation}
	\cancel{G^{\alpha\beta}(\bar{g})+\Lambda_{{\rm cc}} \, \bar{g}^{\alpha\beta} }+ Q^{\alpha\beta}(\bar{g})= \cancel{\kappa^{2} T^{\alpha\beta}(\bar{g}) } \quad \Longrightarrow \quad 
 Q^{\alpha\beta} (\bar{g}) =0,
	\label{method}
\end{equation} 
For the sake of simplicity from now on we will identify $\bar{g}_{ab}$ with $g_{ab}$. 

\subsection{General solutions} 
\label{solution2}
We now investigate under which conditions the G\"{o}del metric is an exact solution of the nonlocal theory (\ref{ac}) for general values of the parameters $m$ and $\omega$. For convenience, we introduce the following truncation of the action (\ref{ac}), 
%
%
\be 
S_n = \int \dd[4]x\sqrt{-g}\, \left[ \frac{1}{2\kappa^2} \left( R-2\Lambda_{{\rm cc}} \right) +\gamma_{0,n}R\square^{n}R+\gamma_{2,n}R_{ab}\square^nR^{ab}+\gamma_{4,n}C_{abcd}\square^nC^{abcd} \right] + S_{m}[g_{ab},\Psi],
\label{sn}
\ee  
where $n$ stays for the $n$-th order of the theory (\ref{ac}).  The nonlocal theory is obtained summing on $n$ from zero to infinity the higher derivative operators. 
%

Let us start with the action $S_0$. It turns out that the theory $S_0$ reduces to Einstein's gravity for 
$\gamma_{i,0}=0$. Therefore, for $\gamma_{i,0}=0 \,\, (i=0,2,4)$, the G\"{o}del metric is an exact solution of the theory.  Solutions for $\gamma_{i,0} \neq 0$ consistent with causality were found in \cite{Accioly:1986wn}.

Next, we study the action $S_n$ for any finite value of the integer $n$.
Making use of (\ref{Riccitensor})-(\ref{dbr}) in (\ref{EoM1}), (\ref{OmegaA}), (\ref{OmegaD}), we can find the 
tensor $Q_{\mu\nu}$ defined in (\ref{Q}) for the local theory (\ref{sn}).
The following result is obtained by mathematical induction for each $n$ for the action $S_n$ with 
$\boxed{n\geq1}$ and assuming $\gamma_{i,0} = 0$, 
\be
\label{order of Q}
&&
\hspace{-1.2cm} 	 Q^{(n)}_{\mu\nu}=\frac{2\kappa^2 }{\Lambda^{2n}}(\gamma_{2,n}+2\gamma_{4,n})(m^2-4\omega^2)2\omega^2(6\omega^2)^{n-1} \times \nonumber \\
	 && \times \mqty(\dmat{(2n+1)m^2-(20+8n)\omega^2,(2n+1)m^2-(12+8n)\omega^2,(2n+1)m^2-(12+8n)\omega^2,4\omega^2-m^2})  .
	 \label{Qn}
\ee
%
The EoM for the nonlocal theory (\ref{actmod}) are obtained taking the sum on the integer $n$ from $n=1$ to $n=+\infty$ of 
 the terms (\ref{Qn}), namely for the moment we assume $\gamma_{i,0} = 0$. The result takes the following compact form after resummation of the form factors, 
\be
&&	\hspace{-0.5cm} 
Q_{\mu\nu}=\sum_{ n=1 }^\infty Q_{\mu\nu}^{(n)}=\nonumber2\kappa^2 \frac{4\omega^2}{\Lambda^2}(m^2-4\omega^2)^2\qty(\gamma'_2(6\omega^2/\Lambda^2)+2\gamma'_4(6\omega^2/\Lambda^2))\mqty(\dmat{1,1,1,0}) \\
&& \hspace{0.5cm}
+2\kappa^2\frac{m^2-4\omega^2}{3}(\gamma_2(6\omega^2/\Lambda^2)+2\gamma_4(6\omega^2/\Lambda^2))\mqty(\dmat{m^2-20\omega^2,m^2-12\omega^2,m^2-12\omega^2,4\omega^2-m^2}) \, , 
	\label{Qinf}
\ee
where $\gamma_i(z)$ is defined in (\ref{fr}) and $\gamma'_i(z)$: 
 \be \gamma'_i(z)=\sum_{i=1}^\infty  \gamma_{i,n} \, n \, z^{n-1} \, , \quad z \equiv \frac{6 \, \omega^2}{\Lambda^2} \, . 
\ee
Finally, the
modified EoM (\ref{dQ}), namely 
\be
G_{\mu\nu}+\Lambda_{\rm cc}g_{\mu\nu}+Q_{\mu\nu}= \kappa^2T_{\mu\nu} \, , 
\ee
for \underline{$\gamma_{i,0} = 0$} read:
\be
&&	
3\omega^2-m^2-\Lambda_{cc}+2\kappa^2\qty[\frac{4\omega^2}{\Lambda^2}(m^2-4\omega^2)^2 \qty(\gamma'_2(z)+2\gamma'_4(z))+\frac{m^2-4\omega^2}{3}(\gamma_2(z)+2\gamma_4(z))(m^2-20\omega^2)]= \kappa^2T_{00}\, ,  \nonumber \\
&& 
\omega^2+\Lambda_{cc}+2\kappa^2\qty[\frac{4\omega^2}{\Lambda^2}(m^2-4\omega^2)^2\qty(\gamma'_2(z)+2\gamma'_4(z))+\frac{m^2-4\omega^2}{3}(\gamma_2(z)+2\gamma_4(z))(m^2-12\omega^2)]=\kappa^2T_{11}= \kappa^2 T_{22} \, , \nonumber \\ 
&& 
m^2-\omega^2+\Lambda_{cc} -2\kappa^2\frac{(m^2-4\omega^2)^2}{3}(\gamma_2(z)+2\gamma_4(z))= \kappa^2 T_{33} \, , 
\ee
where $z\equiv 6\omega^2/\Lambda^2$. 
while the matter content consists on 
the electromagnetic field and a real scalar field \cite{ST} whose total energy-momentum tensor is:
\be 
T_{\mu\nu}= \mqty
(\dmat{\rho+\frac{e^2+E_0^2}{2},p+\frac{E_0^2-e^2}{2},p+\frac{E_0^2-e^2}{2},p+\frac{e^2-E_0^2}{2}}) \, . 
\ee
%
{The parameters $e$ and $E_0$ were introduced in \cite{ST} in order to describe the energy-momentum tensor for a scalar field and for the electromagnetic field respectively , namely 
\be
T_{\mu\nu}^{({\rm Scalar} )}=\text{diag}\left( \frac{e^2}{2},-\frac{e^2}{2},-\frac{e^2}{2},\frac{e^2}{2} \right) \, , \quad 
T_{\mu\nu}^{({\rm EM})}=\text{diag} \left( \frac{E_0^2}{2},\frac{E_0^2}{2},\frac{E_0^2}{2},-\frac{E_0^2}{2} \right)  .
\ee

\underline{So far we assumed $\gamma_{i,0} = 0$, but now we move to consider the general case $\gamma_{i,0}\neq0$. }

According to the explicit calculation of $Q^{(0)}_{\mu\nu}$, which consists in evaluating $P_1^{ab}$, $P_2^{ab}$, and $P_3^{ab}$ in (\ref{EoM1}) but taking only the order $n=0$ in the Taylor's expansion of the form factors,
we get the following non vanishing diagonal contributions to $Q_{\mu\nu}^{(0)}$, 
\be
Q_{00}^{(0)}=2\kappa^2 \qty[\frac{10}{3} \omega ^4 \alpha+m^4 \beta-4 m^2 \omega ^2 \gamma] \, , \nonumber\\
Q_{11}^{(0)}=Q_{22}^{(0)}=2\kappa^2 \qty[2 \omega ^4 \alpha+m^4 \beta-\frac{8}{3} m^2 \omega ^2 \gamma] \, , \nonumber\\
Q_{33}^{(0)}=2\kappa^2 \qty[-\frac{2}{3} \omega ^4 \alpha-m^4\beta+\frac{4}{3} m^2 \omega ^2 \gamma] \, , 
\label{Q0}
\ee
where we have defined the following parameters that take into account of the order zero in the Taylor expansion of the form factors, 
\be
\alpha=3 \gamma_{0,0}+9 \gamma_{2,0}+16 \gamma_{4,0} \, , \nonumber\\
\beta=2 \gamma_{0,0}+\gamma_{2,0}+\frac{2}{3}\gamma_{4,0} \, , \nonumber\\
\gamma=3 \gamma_{0,0}+3 \gamma_{2,0}+4 \gamma_{4,0} \, .
\label{parameters}
\ee

In order to find the full tensor $Q_{\mu\nu}$ for the nonlocal theory we have to sum the contributions $Q_{\mu\nu}^{(n)}$ from $n=0$ to infinity. 
The final result reads:
\be
Q_{\mu\nu} = Q^{(0)}_{\mu\nu} + \sum_{n=1}^{+\infty} Q^{(n)}_{\mu\nu}
=\delta_{\mu\nu} \, \omega^4\qty(a_\mu\frac{m^4}{\omega^4}+b_\mu \frac{m^2}{\omega^2}+c_\mu) 
\label{Qabc}
\ee
where $Q_{\mu\nu}$ is a diagonal matrix and, hence, there is no sum on the index $\mu$. 
In (\ref{Qabc}) the first term is given in (\ref{Q0}) and second term in (\ref{Qinf}).

Therefore, looking at the above expression (\ref{Qabc}) for $Q_{\mu\nu}$ as a polynomial in 
$y \equiv m^2/\omega^2$, one can  easily figure out by linear independence of the monomials $y^0$, $y^1$, and $y^2$ that:
\begin{eqnarray}
\label{gammaneq0}
	 Q_{\mu\nu}=0
	 \,\,\,  \forall \, m,\omega 
	 \quad \Longrightarrow
	 \quad 
	 a_\mu=b_\mu=c_\mu=0 \,\,\,  \forall  \, \omega \quad (\mu= 0,1,2,3) \, . 
	 \label{Qzero}
\end{eqnarray}
Now we rewrite the expressions for $a_\mu$, $b_\mu$, and $c_\mu$ in terms of the form factors. Indeed, after performing the sum over $n$, $a_\mu$, $b_\mu$, and $c_\mu$ will depend on a special linear combination of the form factors $\gamma_2$ and $\gamma_4$, namely $\gamma_2 + 2 \gamma_4$ and its derivative respect to the argument $z=6 \omega^2/\Lambda^2$, i.e. $\gamma^\prime_2 + 2 \gamma^\prime_4$. 
Notice that the form factors in {(\ref{Qinf})}
 do not include the order zero of their Taylor's expansion because we have taken into account of the terms $\gamma_{i,0}$ in $Q^{(0)}_{\mu\nu}$, namely in (\ref{Q0}) or (\ref{parameters}). 
Therefore, the sum on $n$ in $\gamma_2 + 2 \gamma_4$ and $\gamma^\prime_2 + 2 \gamma^\prime_4$ starts from $n=1$. 

The final result for the zero components of the three vectors $a_\mu$, $b_\mu$, and $c_\mu$ reads: 
\be
\label{abc1}
&& 
a_0=\frac{4\omega^2}{\Lambda^2}(\gamma'_2+2\gamma'_4)+\frac{1}{3}(\gamma_2+2\gamma_4)+\beta \, , 
\nonumber\\
&& 
b_0=-\frac{32\omega^2}{\Lambda^2}(\gamma'_2+2\gamma'_4)-8(\gamma_2+2\gamma_4)-4\gamma
\, , 
\nonumber\\
&& 
c_0=\frac{64\omega^2}{\Lambda^2}(\gamma'_2+2\gamma'_4)+\frac{80}{3}(\gamma_2+2\gamma_4)+\frac{10}{3}\alpha \, , 
\ee
while for the first and the second components we have:
\be
\label{abc2}
&&
a_1=a_2=\frac{4\omega^2}{\Lambda^2}(\gamma'_2+2\gamma'_4)+\frac{1}{3}(\gamma_2+2\gamma_4)+\beta
\, , 
\nonumber\\
&&
b_1=b_2=-\frac{32\omega^2}{\Lambda^2}(\gamma'_2+2\gamma'_4)-\frac{16}{3}(\gamma_2+2\gamma_4)-\frac{8}{3}\gamma
\, , 
\nonumber\\
&&
c_1=c_2=\frac{64\omega^2}{\Lambda^2}(\gamma'_2+2\gamma'_4)+16(\gamma_2+2\gamma_4)+2\alpha \, .
\ee
Finally, for the third components we get:
\be
&& \label{abc3}
a_3=-\frac{1}{3}(\gamma_2+2\gamma_4)-\beta
\, , 
\nonumber\\
&&
b_3=\frac{8}{3}(\gamma_2+2\gamma_4)+\frac{4}{3}\gamma
\, , 
\nonumber\\
&&
c_3=-\frac{16}{3}(\gamma_2+2\gamma_4)-\frac{2}{3}\alpha \, . 
\ee
Replacing the expression (\ref{abc3}) into (\ref{abc2}) and imposing (\ref{Qzero}) we get:
\be
  f'(6\omega^2/\Lambda^2)=0  \,\, \forall \, \omega \, , \quad 
\mbox{where } \quad 
f'(6\omega^2/\Lambda^2)=\gamma'_2(6\omega^2/\Lambda^2)+2\gamma'_4(6\omega^2/\Lambda^2) \, . 
\label{above}
\ee
Simplifying the equations $a_3=b_3=c_3=0$ taking into account of (\ref{parameters}) 
we get:
\be
f(6\omega^2/\Lambda^2)+\gamma_{2,0}+2\gamma_{4,0}
= \sum_{n=1}^{+\infty} (\gamma_{2,n} + 2 \gamma_{4,n} )\qty(\frac{6\omega^2}{\Lambda^2})^n + \gamma_{2,0}+2\gamma_{4,0}
=0 
 \,\, \, \forall\omega \, , 
 \qand 3\gamma_{0,0}+\gamma_{2,0}=0
\, .
\label{solz}
\ee
Including the constant term again in the sum we can rewrite (\ref{solz}) as follows,
\be
 \sum_{n=0}^{+\infty} (\gamma_{2,n} + 2 \gamma_{4,n} )\qty(\frac{6\omega^2}{\Lambda^2})^n \equiv 
 \gamma_{2} + 2 \gamma_{4} 
=0 
 \,\, \, \forall\omega \, , 
 \qand 3\gamma_{0,0}+\gamma_{2,0}=0
\, .
\label{solz2}
\ee
Hence, the final result for the nonlocal theory is:
\begin{eqnarray}
	\boxed{Q_{\mu\nu}=0 \,\,\, \forall \, m,\omega\quad \Longleftrightarrow \quad \gamma_2+2\gamma_4=0\text{ (sum from 0 to infinity)}\qand  3\gamma_{0,0}+\gamma_{2,0}=0} \, .
	\label{condNL}
\end{eqnarray}

For a general local theory defined by the action,
\be 
S_N = \int \dd[4]x\sqrt{-g}\, \left[ \frac{1}{2\kappa^2}\qty(R-2\Lambda_{\rm{cc}})+
\sum_{n=0}^N \gamma_{0,n}R\square^{n}R+ \sum_{n=0}^N \gamma_{2,n}R_{ab}\square^nR^{ab}+\sum_{n=0}^N \gamma_{4,n}C_{abcd}\square^nC^{abcd} \right] + S_{m},
\label{snN}
\ee  
the condition (\ref{condNL}) holds whether we include the extra condition $\gamma_{i,n}=0$ for $n>N$, namely the condition (\ref{condNL}) applies to the coefficients $\gamma_{i,n}$ for $0\leqslant n \leqslant N$, i.e. 
\begin{eqnarray}
	\boxed{Q_{\mu\nu}=0 \,\,\, \forall \, m,\omega\quad \Longleftrightarrow \quad (\gamma_{2,n}+2\gamma_{4,n})=0 \quad {\rm for} \quad 0\leqslant n \leqslant N  \qand  3\gamma_{0,0}+\gamma_{2,0}=0} \, .
	\label{condL}
\end{eqnarray}




%

\subsection{Exact solutions without CTC in Nonlocal Gravity}
\label{solutionsNoCTC}
Another class of particular G\"{o}del exact solutions for local as well as nonlocal gravitational theories is obtained for $m^2=4\omega^2$ \cite{Nascimento:2021bzb} 
regardless of the explicit form of the form factors as long as the constant term in the form factors $\gamma_i$ is zero, namely $\gamma_{i,0} = 0$.  Indeed, both $Q^{(n)}_{\mu\nu}$ and $Q_{\mu\nu}$ are identically zero for $m^2-4\omega^2=0$ whether $\gamma_{i,0} = 0$ (see (\ref{Qn}) and (\ref{Qinf})). 

However, if the conditions (\ref{condNL}) and (\ref{condL}) on the nonlocal or local form factors are satisfied, 
then the rotating spacetimes for which $m^2=4\omega^2$ are exact solutions although $\gamma_{i,0} \neq0$. 

It deserves to be notice that according to (\ref{CTCcond}) for $m^2=4\omega^2$ there are no CTC. Therefore, rotating causal Universes are exact solutions of nonlocal and local theories. 
%

%

\subsection{A more general action}
\label{GNoCTC}
We can extend the class of solutions with $m^2-4\omega^2=0$ found in the previous section to a more general class of theories. 
Let us consider the action (\ref{ac}) augmented by other operators whose prototype is the following one,
\begin{equation}
	\label{1d}
	\mathcal{L}_{a,b}=A_{a_1\cdots a_{n}}\nabla^{a_{n}}B^{a_1\cdots a_{n-1}},
\end{equation}
where
$A_{a_1\cdots a_{n}}$ and $B_{a_1\cdots a_{n-1}}$ are general tensors made of any finite number of derivatives (including no derivatives) and one or more curvature tensors 
(for example: $A_{abcd}=(\nabla^{f}R)R_a{}^eR_{eb}\nabla_f R_{cd}$).  
%
Now performing an explicit computation, it turns out that
\begin{equation}
	\label{dr}
	\nabla_a R_{bcde}\propto (m^2-4\omega^2) \, ,  \quad {\mbox{or in short}}: \quad 
	\nabla \, {\rm Riem}  \propto (m^2-4\omega^2) \, , 
\end{equation}
which vanishes for $m^2-4\omega ^2=0$. Taking into account of (\ref{dr}) 
the variation of the action operator for the Lagrangian term (\ref{1d}) with respect to the metric gives: 
	\be
	\label{vo1d}
	&&	\delta \!\! \int\dd[4]{x}\sqrt{-g}\mathcal{L}_{a,b}=\int\dd[4]{x}\sqrt{-g}\bigg[\mathcal{L}_{a,b}\frac{g^{ab}}{2} \delta g_{ab}+\delta(A_{a_1\cdots a_{n}})\nabla^{a_{n}}B^{a_1\cdots a_{n-1}}+ \nonumber \\
	&&\hspace{2.5cm}
	 \qquad A_{a_1\cdots a_{n}}\delta(\nabla^{a_{n}})B^{a_1\cdots a_{n-1}}+A_{a_1\cdots a_{n}}\nabla^{a_{n}}\delta(B^{a_1\cdots a_{n-1}})\bigg] \, . 
	\ee
	In the above variation the first and the second term contain $\nabla \,  {\rm Riem}$. Therefore, according to (\ref{dr}) they are zero when evaluated for $m^2 = 4 \omega^2$. So far the variation reads:
	\be
	 \delta \!\!  \int\dd[4]{x}\sqrt{-g}\mathcal{L}_{a,b}\Big|_{m^2 = 4 \omega^2} 
	 =\int\dd[4]{x}\sqrt{-g}\qty[A_{a_1\cdots a_{n}}\delta(\nabla^{a_{n}})B^{a_1\cdots a_{n-1}}-\delta(B^{a_1\cdots a_{n-1}})\nabla^{a_{n}}A_{a_1\cdots a_{n}}] \, , 
	 \ee
	 where the second term resulting from the integration by parts is again zero because of (\ref{dr}). 
	 Let us now compute the variation of the covariant derivative,
	 \be
	 \delta \!\! \int\dd[4]{x}\sqrt{-g}\mathcal{L}_{ a,b}\Big|_{m^2 = 4 \omega^2} 
	&=&\int\dd[4]{x}\sqrt{-g}\qty[A_{a_1\cdots a_{n}}\delta(g^{a_na})\nabla_{a}B^{a_1\cdots a_{n-1}}+A_{a_1\cdots a_{n-1}}{}^{a}\delta(\nabla_{a})B^{a_1\cdots a_{n-1}}] \nonumber \\
	%
	&=& \int\dd[4]{x}\sqrt{-g}\qty[A_{a_1\cdots a_{n-1}}{}^{a}\sum_{i=1}^{n-1}\qty(\delta\Gamma^{a_i}{}_{a b})B^{a_1\cdots a_{i-1}ba_{i+1}\cdots a_{n-1}}] \, , 
	\label{finaldelta}
\ee
where the variation of the connection is:
\be
\delta\Gamma^{c}{}_{ab}=\frac{1}{2}\qty(\nabla_a h_{b}{}^c+\nabla_b h_{a}{}^c-\nabla^c h_{ab})
\, , \quad \delta g_{ab}\equiv h_{ab} \, .
\ee
Integrating by parts the derivatives present in the variation of the connection, we end up with expression containing derivatives of the tensors $A$ and $B$. Hence, the variation (\ref{finaldelta}) is zero 
among using one more time (\ref{dr}) for $m^2 = 4 \omega^2$.

Except for the last step of Eq. (\ref{vo1d}), the other terms vanish because these terms contain $\nabla_a R_{bcde}$. Such terms will vanish when $m^2-4\omega ^2=0$ due to (\ref{dr}) and integration by parts in the last step of Eq. (\ref{vo1d}) will introduce the derivative operator in $A_{a_1\cdots a_{n-1}}{}^{a}B^{a_1\cdots a_{i-1}ba_{i+1}\cdots a_{n-1}}$, so the result is zero when $m^2-4\omega ^2=0$.
%
\begin{mdframed}
We can conclude that if a general Lagrangian contains operators with at least one derivative, besides the Einstein-Hilbert term in presence of cosmological constant, then, the metrics for which $m^2-4\omega ^2=0$ (without CTC) are exact solutions of the theory. 
\end{mdframed}
The above statement include the result at the end of the previous section relative to the case $m^2 - 4 \omega^2=0$. Indeed, the G\"{o}del metric with $m^2 = 4 \omega^2$ is a solution in nonlocal gravity if all the constant terms in the Taylor expansion of the form factors 
$\gamma_{i,0}$ are zero regardless of the explicit form of the form factors. 


\section{
Quantum renormalizablity and CTC}
\label{quantum}
%
We here investigate the presence of CTC in a special class of classical nonlocal theories compatible with unitary and (super-)renormalizability. Indeed, in the previous section we did not assume any relation between the form factors $\gamma_0$, $\gamma_2$, and $\gamma_4$ and we found that the G\"{o}del spacetimes exact solutions if the conditions (\ref{condNL}) are satisfied by the theory.
In this section we compare the result (\ref{condNL}) with the relations that the form factors should satisfy in order to have theories  compatible with unitary and renormalizability. 

The latter properties had been extensively studied in letterature for the following Lagrangian \cite{Koshelev:2016xqb} written in the Ricci-Riemann bases, 
\be
&& \mathcal{L} = \frac{1}{2\kappa^2} \left( R-2\Lambda_{{\rm cc}}  \right)  +R\tilde\gamma_{0}(\square)R+R_{ab}\tilde\gamma_{2}(\square)R^{ab}+R_{abcd}\tilde\gamma_{4}(\square)R^{abcd}  \, , 
\label{tildeF}\\
&& \nonumber \\
&& \tilde\gamma_0(\Box) =-\frac{(D-2)\qty(e^{\mathrm H_0(\Box)}-1)+D\qty(e^{\mathrm H_2(\Box)}-1)}{ 2\kappa^2 \, 4(D-1)\Box}+ \tilde\gamma_4(\Box) \, , \label{gamma0} \\
&& \nonumber \\
&& \tilde\gamma_2(  \Box)=\frac{e^{\mathrm H_2(\Box)}-1}{2 \kappa^2  \, \Box}-4\tilde\gamma_4(\Box) \, , 
\label{gamma2p}
\ee
where the form factors have been properly selected in order to end up with the most general propagator (we here remind only the gauge invariant parto of the propagator) consistent with unitarity and (super-)renormalizablity \cite{Koshelev:2017ebj, Koshelev:2016xqb}, namely 
%
\be
\mathcal{O}^{-1}(k)=-\frac{1}{k^2}\qty[\frac{P^{(2)}}{e^{\mathrm H_2(k^2)}}-\frac{P^{(0)}}{(D-2)e^{\mathrm H_0(k^2)}}] \, , 
\ee
where $\{P^{(i)}|\; i=0,2\}$ are the projectors \cite{modestoDdim}. $\mathrm H_0$ and $\mathrm H_2$ are non-zero entire functions 
asymptotically approaching the same logarithm of a polynomial in the variable $k^2$ (at least in the simplest version of the theory). 
The entire functions $H_2$ and $H_0$ must have the same asymptotic behaviour in order to achieve renormalizabilty, while according to tree-level \cite{Briscese:2018bny, Briscese:2019rii, QGscattering, Tree-LevelScattering} and perturbative Unitarity \cite{Briscese:2018oyx, Briscese:2021mob, Liu:2022gun} $H_2(0) = H_0(0) = 0$.

The G\"{o}del solution has been studied for the theory in the Ricci-Weyl basis (\ref{ac}), thus, in order to infer about a possible relation with unitarity and renormalizabilty we have to change basis. 
In the appendix (\ref{RB}) we derived the relation between the form factors $\tilde\gamma_i$ (\ref{tildeF}) and the form factors $\gamma_i$ (\ref{ac}). The outcome is:
\begin{equation}
	\label{rienbasis2}
	\gamma_2+2\gamma_4 =\tilde\gamma_2+4\tilde\gamma_4  \, .
\end{equation}
Therefore, according to (\ref{condNL}) the metrics for general $m$ and $\omega$ are exact solutions of the theory (\ref{tildeF}) if:
\begin{equation}
	\label{rienbasis3}
	\tilde\gamma_2 = - 4\tilde\gamma_4  \, .
\end{equation}
Replacing the above identity in (\ref{gamma2p}) we get: $H_2 = 0$, which is inconsistent with the renormalizability of the theory. 
Hence we can make the following statement,

\begin{mdframed}
\mbox{G\"{o}del spacetimes with CTC are not exact solutions in nonlocal (super-)renormalizable gravitational theories. }
\end{mdframed}

 In other words, super-renormalizability and the G\"{o}del's spacetimes with CTC are incompatible.
 Notice that the Unitarity is consistent with the G\"{o}del's spacetimes with CTC because $H_2=0$ does not change the residue at the Cutkosky cuts \cite{Briscese:2018oyx, Liu:2022gun}.
 
For the case a local theories, one have to replace $\exp H_2$ with a polynomial that must be zero for consistency with (\ref{rienbasis3})) whether we want the G\"{o}del's spacetimes to be solutions. Hence, the local theories that predict  spacetimes with CTC are non-renormalizable. 
 
According to section \ref{solutionsNoCTC}, it is doubly surprising that metrics with CTC are not solutions while those without CTC are so in practically all theories.

 \section{Nonlocal gravity non-minimally coupled to matter} \label{nm-gravity}
In this section, we very briefly review the nonlocal gravitational theory coupled to matter proposed in \cite{nl-nm-matter}. 
The classical action is: 
\be
&&
\label{actionNM}
S [\Phi] = \int d^D x \sqrt{|g|} \left[ \mathcal{L}_{\rm loc} + E_i F^{ij} (\hat{\Delta}) E_j \right]
\, , 
\\
&&
S_{\rm loc} = \int d^D x \sqrt{|g|} \, \mathcal{L}_{\rm loc}
\, , \quad 
\mathcal{L}_{\rm loc} = \frac{1}{2 \kappa^2} R  + \mathcal{L}_{\rm m} (g_{\mu\nu}, \phi, \psi, A^\mu)
\, , 
\label{Lloc}
\\
&&
E_i (x) = \frac{\delta S_{\rm loc}}{\delta \Phi^i (x)}
\, , \quad 
\Delta_{ij} (x,y) = \frac{\delta E_{i} (x)}{\delta \Phi^j(y)} =  \frac{\delta^2 S_{\rm loc}}{\delta \Phi^j(y) \delta \Phi^i (x)} = \hat{\Delta}_{ij} \, \delta (x,y) \, , \quad 
( \hat{\Delta}_{\Lambda}) _{ij} = \frac{\hat{\Delta}_{ij}}{\left( \Lambda\right)^{[\hat{\Delta}_{ij}]}} \, 
,
\\
&&
2 \hat{\Delta}_{ik} \, F^{k} {}_j (\hat{\Delta}) = \left[ e^{{\rm H}(\hat{\Delta}_{\Lambda} )}   - 1 \right]_{ij} \, ,   \label{FF2}
\ee
where 
 by $\Phi^i \equiv (g_{\mu\nu}, \phi, \psi, A^\mu)$ we mean any field, $F^{ij}$ is a symmetric tensorial entire function whose argument is the Hessian operator $\hat{\Delta}_{ij}$, and ${\rm H}( \hat{\Delta}_{\Lambda_*} )$ is an entire analytic function whose argument is the dimensionless Hessian. 
In the above formula, we used 
the notation $[X]$ to indicate the dimensionality of the quantity $X$ in powers of mass units, i.e., $X$ has dimension of $(\text{mass})^{[X]}$. Since $[\Lambda]=1$, it follows that $[( \hat{\Delta}_{\Lambda}) _{ij}]=0$, as claimed.

 The theory (\ref{actionNM}), enjoys the following essential properties. 
(i) All the solutions of Einstein's gravity coupled to matter, namely the solutions of the local theory (\ref{Lloc}), are solutions of the nonlocal theory too. Indeed, 
the equations of motion of (\ref{actionNM}) read:
\be
\left[e^{{\rm H}(\hat{\Delta}_{\Lambda_*})}\right]_{kj} \, E_j + O(E^2) = 0 \, , \label{LEOM}
\ee
where $E_i$ are the EoM of the local theory. 
(ii) 
The nonlocal theory gives the same tree-level scattering amplitudes of the local theory \cite{QGscattering, Tree-LevelScattering}. This latter property guarantees macro-causality \cite{CausalityGiaccari}, namely the Shapiro's time delay evaluated in the eikonal approximation is the same of the one in Einstein's local theory coupled to the standard model of particle physics.
(iii) 
The stability properties are the same at linear and non linear level of the local theory whether we perturb an exact solution of the local theory \cite{Briscese:2019rii, Briscese:2018bny}.
(iv) 
The theory is super-renormalizable or finite at quantum level \cite{kuzmin, modesto, modestoDdim}, and unitary at any perturbative order in the loop expansion \cite{Briscese:2019rii}. 
%

Among the above four properties the most important one for what concerns the topic of this paper is the the first one. 
Indeed, if we include in the local theory a cosmological term, all the G\"{o}del-type metrics are exact solutions of the theory (\ref{actionNM}). In particular, the G\"{o}del-type spacetimes with CTC turn out to be solutions of (\ref{actionNM}). However, the quantum effective action will include perturbative corrections to the form factor that very likely will not satisfy the condition (\ref{rienbasis3}). Therefore, the G\"{o}del-type metrics with CTC will not be solution of the quantum effective equations of motion.

\section*{Conclusions}
We have investigated whether the homogeneous G\"{o}del-type metrics can be exact solutions of a general classe of nonlocal and local gravitational theories minimally coupled to matter. 
 
It tuned out that the G\"{o}del's metrics without CTC are basically solutions of all nonlocal as well as local higher derivatives theories, while the G\"{o}del's spacetimes with CTC that solve the Einstein's theory do not solve the EoM of (super-)renormalizable gravitational theories. It turns out that the super-renormalizability property is the guiding to select out theories consistent with the G\"{o}del-type causality. Indeed, we showed that unitarity alone is not enough to guarantee such kind of cosmological causality. In particular, the spacetimes with CTC are exact solutions of a large class of nonlocal ghost-free but non-renormalizable theories.

In another class of nonlocal gravitational theories with non-minimal coupling to matter, all the G\"{o}del's spacetimes are exact solutions of the classical theory, and, thus the causality violation is manifest.
However, very likely the quantum corrections will spoil the above statement that needs a very special relation between the $R^2$'s and the ${\rm Ric}^2$'s quantum form factors, namely the relation between the following two operators,
\be
R f_0(\Box) R \quad {\rm and} \quad {\rm Ric} f_2(\Box) {\rm Ric} .
\ee

Therefore, we are entitled to state that (super-)renormalizabily and G\"{o}del's causality violation exclude each other.

\section*{Acknowledgement}
This work was supported by the Basic Research Program of the Science, Technology, and Innovation Commission of Shenzhen Municipality (grant no. JCYJ20180302174206969). 

\appendix
\label{appendix}


\section{Equations of motion for local and analytic nonlocal theories}
As a general operator, we consider the following general action:
\be
S_{a,m}=\int\dd[4]{x}\sqrt{-g} \, \mathcal{L}_{a,m}=\int\dd[4]{x}\sqrt{-g} \, \gamma_{a,m}A_{a_1\cdots a_{n}}\Box^mB^{a_1\cdots a_{n}},
\ee
where $\gamma_{a,m}$ are the coefficients of the power series for the form factors. 

Taking the variation of the action respect to the metric, we get:
\begin{equation}
	\begin{aligned}
	\delta \! 
	\int\dd[4]{x}\sqrt{-g}A_{a_1\cdots a_{n}}\Box^mB^{a_1\cdots a_{n}}&=\int\dd[4]{x}\sqrt{-g}\qty[\frac{g_{ab}}{2}A_{a_1\cdots a_{n}}\Box^mB^{a_1\cdots a_{n}}h_{ab}+\delta\qty(A_{a_1\cdots a_{n}}\Box^mB^{a_1\cdots a_{n}})]\\&=\int\dd[4]{x}\sqrt{-g}\bigg[\frac{g_{ab}}{2}A_{a_1\cdots a_{n}}\Box^mB^{a_1\cdots a_{n}}h_{ab}+\delta\qty(A_{a_1\cdots b_{n}})\Box^mB^{a_1\cdots a_{n}}\\&\qquad+A_{a_1\cdots a_{n}}\delta\qty(\Box^m)B^{a_1\cdots a_{n}}+A_{a_1\cdots a_{n}}\Box^m\delta\qty(B^{a_1\cdots a_{n}})\bigg]\\&=\int\dd[4]{x}\sqrt{-g}\bigg[\frac{g_{ab}}{2}A_{a_1\cdots a_{n}}\Box^mB^{a_1\cdots a_{n}}h_{ab}+\delta\qty(A_{a_1\cdots a_{n}})\Box^mB^{a_1\cdots a_{n}}\\&\qquad+\sum_{i=0}^{m-1}(\Box^iA_{a_1\cdots a_{n}})\delta\qty(\Box)\qty(\Box^{m-1-i}B^{a_1\cdots a_{n}})+\delta\qty(B^{a_1\cdots a_{n}})\Box^mA_{a_1\cdots a_{n}}\bigg] .
	\end{aligned}
	\label{vga}
\end{equation}

Now we list some useful formulas:
\be 
&& \Box(AB)=(\Box A)B+2(\nabla_aA)(\nabla^a B)+A(\Box B), \\
&&
	(\delta\Box)T^{\cdots}{}_{\cdots}=\delta(g^{ab}\nabla_a\nabla_b)T^{\cdots}{}_{\cdots}=-(\nabla^a\nabla^bT^{\cdots}{}_{\cdots})h_{ab}+\qty[\sum_{}\delta\Gamma^{.}{}_{..} (\nabla_aT^{\cdots}{}_{\cdots})+\sum_{}\nabla_a(\delta\Gamma^{.}{}_{..} T^{\cdots}{}_{\cdots})]g^{ab},
	\label{db}
\ee
where the variation of the connection respect to the metric reads:
\be
\delta\Gamma^{c}{}_{ab}=\frac{1}{2}\qty(\nabla_a h_{b}{}^c+\nabla_b h_{a}{}^c-\nabla^c h_{ab}).
\ee
Using the formulas of above we can find the EoM for the theory (\ref{actmod}), namely
%
\begin{eqnarray}
\nonumber 
&& 
\hspace{-1cm}
E^{ab}=G^{ab}+\Lambda_{\mbox{cc}} g^{ab}+P_{1}^{ab}+P_{2}^{ab}+P_{3}^{ab}-2\Omega_{1}^{ab}+g^{ab}(g_{cd}\Omega^{cd}_{1}+\bar{\Omega}_{1})-2\Omega^{ab}_{2}\\
&& + g^{ab}(g_{cd}\Omega^{cd}_{2}+\bar{\Omega}_{2})-4\Delta^{ab}_{2}-2\Omega^{ab}_{3}+g^{ab}(g_{cd}\Omega^{cd}_{3}+\bar{\Omega}_{3})-8\Delta^{ab}_{3} 
- \kappa^{2}T^{ab} = 0 \, ,
\label{ac1}
\end{eqnarray} 
where the tensors $P_{i}^{ab}$ in (\ref{ac1}) are defined as follows, 
 \begin{eqnarray}
\nonumber
 P_{1}^{ab}&=&\kappa^2 \left[\left(4G^{ab}+g^{ab}R-4(\nabla^{a}\nabla^{b}-g^{ab}\square)\right)\gamma_0(\square)R\right] \, , \\
\nonumber P_{2}^{ab}&=&\kappa^2\bigg[4R^{{(}a}_{\,d}\gamma_2(\square)R^{{|}d{|}b{)}}-g^{ab}R^{cd}\gamma_2(\square)R_{cd}-4\nabla_{d}\nabla^{{(}b}(\gamma_2(\square)R^{{|}d{|}a{)}})+ 2\square(\gamma_2(\square)R^{ab})+2g^{ab}\nabla_{c}\nabla_{d}(\gamma_2(\square)R^{cd})\bigg] \, , \\
 P_{3}^{ab}&=&\kappa^2\bigg[-g^{ab}C^{cdef}\gamma_4(\square)C_{cdef}+4C^{{(}a}_{\,\,cde}\gamma_4(\square)C^{b{)}cde} -4(R_{cd}+2\nabla_{c}\nabla_{d})(\gamma_4(\square)C^{{(}b{|}cd{|}a{)}})\bigg],
\label{EoM1}
\end{eqnarray}
while the tensors $\Omega^{ab}_{i}$ and $\tilde\Omega^{ab}_{i}$ in (\ref{ac1}) read:
\begin{eqnarray}	
\nonumber 
\Omega^{ab}_{1}&=&\kappa^2\sum_{n=1}^{\infty}\gamma_{0,n}{\frac{1}{\Lambda^{2n}}}\sum_{l=0}^{n-1}\nabla^{a}R^{(l)}\nabla^{b}R^{(n-l-1)}, \quad \bar{\Omega}_{1}=\kappa^2\sum_{n=1}^{\infty}\gamma_{0,n}{\frac{1}{\Lambda^{2n}}}\sum_{l=0}^{n-1}R^{(l)}R^{(n-l)} \, , \\
\nonumber\Omega^{ab}_{2}&=&\kappa^2\sum_{n=1}^{\infty}\gamma_{2,n}{\frac{1}{\Lambda^{2n}}}\sum_{l=0}^{n-1}(\nabla^{a}R^{cd(l)})(\nabla^{b}R_{cd}^{(n-l-1)}),\quad \bar{\Omega}_{2}=\kappa^2\sum_{n=1}^{\infty}\gamma_{2,n}{\frac{1}{\Lambda^{2n}}}\sum_{l=0}^{n-1}R^{cd(l)}R_{cd}^{(n-l)} \, , \\
\nonumber\Omega^{ab}_{3}&=&\kappa^2\sum_{n=1}^{\infty}\gamma_{4,n}{\frac{1}{\Lambda^{2n}}}\sum_{l=0}^{n-1}(\nabla^{a}C^{c(l)}_{\,\,def})(\nabla^{b}C_{c}^{\,\,def(n-l-1)}) \, \\
 \tilde{\Omega}_{3}&=&\kappa^2\sum_{n=1}^{\infty}\gamma_{4,n}{\frac{1}{\Lambda^{2n}}}\sum_{l=0}^{n-1}C^{a(l)}_{\,bcd}C^{\,\,bcd(n-l)}_{a}.
\label{OmegaA}
\ee
Finally, $\Delta^{ab}_{i}$ are:
\be
\nonumber\Delta^{ab}_{2}&=&{\kappa^2}\sum_{n=1}^{\infty}\gamma_{2,n}{\frac{1}{\Lambda^{2n}}}\sum_{l=0}^{n-1}\nabla^{c}\left(R_{dc}^{(l)}\nabla^{(a}R^{b)d(n-l-1)}-(\nabla^{(a}R_{dc})R^{b)d(n-l-1)}\right) \, , \\
\Delta^{ab}_{3}&=&{\kappa^2}\sum_{n=1}^{\infty}\gamma_{4,n}{\frac{1}{\Lambda^{2n}}}\sum_{l=0}^{n-1}\nabla^{c}\left(C_{\,\,cef}^{d(l)}\nabla^{(a}C^{\,\,b)ef(n-l-1)}_{d}-(\nabla^{(a}C_{\,\,cef}^{|d(l)|})C^{b)ef(n-l-1)}_{d}\right) ,
\label{OmegaD}
\end{eqnarray}
here we are used the notation 
$A^{(l)}\equiv \square^{l}A$.

In order to simplify the EoM, it is convenient to rewrite the homogeneous G\"{o}del-type metrics in the 
orthonormal-tetrad formalism,
\be 
\label{basis-g} g_{ab}=\eta_{\mu\nu}(e^\mu)_a(e^\nu)_b\qand  \{(e^{\mu})_a\} = {\rm diag}  \left( \dd{t}+H(r)\dd{\theta}, \dd r,D(r)\dd\theta,\dd z  \right) \, .
\label{tetraG}
\ee 
Now we are ready to replace the ansatz (\ref{tetraG}) in the EoM. 

According to (\ref{tetraG}), an explicit but tedious computation gives:
\be
\label{Riccitensor} 
R_{\mu\nu}=\mqty(\dmat{2\omega^2,2\omega^2-m^2,2\omega^2-m^2,0}) \qand R=2(\omega^2-m^2) \, .
\ee 
Therefore, the Einstein's tensor reads:
\be
\label{Einsteintensor} 
 G_{\mu\nu}=\mqty(\dmat{3\omega^2-m^2,\omega^2,\omega^2,m^2-\omega^2}) \, . 
\ee 
In the same vain, we evaluate other invariant tensors present in the EoM, namely 
\be 
&& R^{ab}R_{ab}=2(m^4-4m^2\omega^2+6\omega^4)\, ,  \nonumber \\
&&
R^{ab}\Box R_{ab}=4\omega^2(4\omega^2-m^2)^2 \, , \nonumber \\
&& 
 C^{abcd}C_{abcd}=\frac{4}{3}(m^2-4\omega^2)^2 \qand C^{abcd}\Box C_{abcd}=8\omega^2(m^2-4\omega^2)^2 \, , \nonumber \\
 &&
\grad_a\grad_b\Box^n R^{ab}=0\qand \grad_a\Box^n R^{ab}=0\qfor n= 0,1 \, . 
 \ee
Finally, a very useful formula is:
\be
\label{useful}  \Box^2R_{abcd}=6\omega^2\Box R_{abcd} \, , 
\ee
which shows the equivalence of acting with higher derivatives on the Riemann's tensor and the multiplication by $6\omega^2$.
 Hence, according to (\ref{useful}), we easily get:
\be 
&& R^{ab}\Box^n R_{ab}=\frac{2}{3}(6\omega^2)^n (4\omega^2-m^2)^2\qand C^{abcd}\Box^n C_{abcd}=\frac{4}{3}(6\omega^2)^n (m^2-4\omega^2)^2\qfor n \geq1 \, , \nonumber \\
&&
\label{dbr} \grad_a\grad_b\Box^n R^{ab}=0\qand \grad_a\Box^n R^{ab}=0\qfor n \geq0 \, . 
 \ee

\section{Nonlocal gravity in the Riemann-Ricci basis}
\label{RB}

In this section we derive the relation betwee the theory in the Weyl-Ricci basis (\ref{ac}) to the theory in the Riemann-Ricci basis (\ref{tildeF}). 
We start by recalling the following definition of the Weyl tensor in dimension $D$,
\be
C_{abcd}=R_{abcd}-\frac{2}{D-2}\qty(g_{a[c}R_{d]b}-g_{b[c}R_{d]a})+\frac{2}{(D-1)(D-2)}Rg_{a[c}g_{d]b} \, ,  
\ee
(notice that $C_{abcd}$ is traceless). Afterwards, we evaluate the Weyl square scalar, namely 
\be 
C_{abcd}C^{abcd}=R_{abcd}C^{abcd}=C_{abcd}R^{abcd}=\frac{2}{(D-1)(D-2)}R^2-\frac{4}{D-2}R_{ab}R^{ab}+R_{abcd}R^{abcd} \, .
\ee
Since $\nabla_cg_{ab}=0$ and $C_{abcd}$ is a linear function of the Riemann tensor $R_{abcd}$, we have 
	\be
	&&	\hspace{-0.35cm}
	\mathcal{L}_g = \frac{R-2\Lambda_{\rm{cc}}}{2\kappa^2} +R\gamma_{0}(\square)R+R_{ab}\gamma_{2}(\square)R^{ab}+C_{abcd}\gamma_{4}(\square)C^{abcd} \nonumber \\
		&& = \frac{R-2\Lambda_{\rm{cc}}}{2\kappa^2}   +R\gamma_{0}(\square)R+R_{ab}\gamma_{2}(\square)R^{ab}+\frac{2}{(D-1)(D-2)}R\gamma_{4}(\square)R-\frac{4}{D-2}R_{ab}\gamma_{4}(\square)R^{ab}+R_{abcd}\gamma_{4}(\square)R^{abcd} \nonumber \\
		&&
		= \frac{R-2\Lambda_{\rm{cc}}}{2\kappa^2} +R\qty[\gamma_{0}(\square)+\frac{2}{(D-1)(D-2)}\gamma_{4}(\square)]R+R_{ab}\qty[\gamma_{2}(\square)-\frac{4}{D-2}\gamma_{4}(\square)]R^{ab}+R_{abcd}\gamma_{4}(\square)R^{abcd} \, , 
		\label{redef}
		\ee
		which has to be equal to (\ref{tildeF}), namely 
		\be
		\frac{1}{2\kappa^2} \left( R-2\Lambda_{\rm{cc}} \right) +R\tilde\gamma_{0}(\square)R+R_{ab}\tilde\gamma_{2}(\square)R^{ab}+R_{abcd}\tilde\gamma_{4}(\square)R^{abcd} \, .
		\label{tildeF2}
	\ee
Comparing the last step in (\ref{redef}) with the Lagrangian (\ref{tildeF2}), 
\be
\tilde\gamma_0=\gamma_0+\frac{2}{(D-1)(D-2)}\gamma_4\, , \quad \tilde\gamma_2=\gamma_2-\frac{4}{D-2}\gamma_4 \, , \qand \tilde\gamma_4=\gamma_4 \, . 
\ee
For $D=4$, we have the following relation between form factors $\gamma_i$ and $\tilde{\gamma}_i$, 
\begin{equation}
	\label{rienbasis}
	\gamma_2+2\gamma_4=(\gamma_2-2\gamma_4)+4\gamma_4=\tilde\gamma_2+4\tilde\gamma_4  \, .
\end{equation}

\end{document}